\UseRawInputEncoding
\documentclass{article}
\usepackage{spconf,amsmath,graphicx}
\usepackage{cite}
\usepackage{amssymb}
\usepackage{multirow}
\usepackage{booktabs}
\usepackage{color}
\usepackage{caption}

\title{A noise-robust self-supervised pre-training model based speech representation learning for automatic speech recognition}
\name{Qiu-Shi Zhu$^1$, Jie Zhang$^{1,2}$, Zi-Qiang Zhang$^1$, Ming-Hui Wu$^1$, Xin Fang$^1$, Li-Rong Dai$^1$
\thanks{This work was supported by the National Natural Science Foundation of China (No. 62101523),
Fundamental Research Funds for the Central Universities and the Leading Plan of CAS (XDC08010200).}}
\address{ $^1$NEL-SLIP, University of Science and Technology of China (USTC), Hefei, China\\
  $^2$State Key Laboratory of Acoustics, Institute of Acoustics, Chinese Academy of Sciences, Beijing, China}

\begin{document}
\ninept
\maketitle
\begin{abstract}
Wav2vec2.0 is a popular self-supervised pre-training framework for learning speech representations in the context of automatic speech recognition (ASR).
It was shown that wav2vec2.0 has a good robustness against the domain shift, while the noise robustness is still unclear.
In this work, we therefore first analyze the noise robustness of wav2vec2.0 via experiments.
We observe that wav2vec2.0 pre-trained on noisy data can obtain good representations and thus improve the ASR performance on the noisy test set, which however brings a performance degradation  on the clean test set.
To avoid this issue, in this work we propose an enhanced wav2vec2.0 model.
Specifically, the noisy speech and the corresponding clean version are fed into the same feature encoder, where the clean speech provides training targets for the model.
Experimental results reveal that the proposed method can not only improve the ASR performance on the noisy test set which surpasses the original wav2vec2.0,  but also ensure a tiny  performance decrease on the clean test set.
In addition, the effectiveness of the proposed method is demonstrated under different types of noise conditions.

\end{abstract}
\section{Introduction}
\label{sec:intro}

Self-supervised pre-training has become an effective method for neural network models to utilize unlabeled data.
Recently, many self-supervised learning methods for speech representations have been proposed in the speech domain.
For example, autoregressive predictive coding (APC) \cite{Chung2019} was proposed to reconstruct the future frames based on the past frames.
Contrastive predictive coding (CPC) \cite{oord2018representation} and wav2vec \cite{Schneider2019} perform the next-step prediction in a similar way but using a contrastive loss.
Meanwhile, the contextual speech representations can be learned from the unlabeled speech data by reconstructing the masked input speech frames in~\cite{liu2020mockingjay,liu2021tera,jiang2019improving,ling2020deep,wang2020unsupervised}.
Specifically, the contextual speech representations learned in ~\cite{liu2020mockingjay,liu2021tera,jiang2019improving} utilize a bidirectional transformer (BERT) structure \cite{devlin2018bert}, while that learned in~\cite{ling2020deep, wang2020unsupervised} are based on a bidirectional long short-term memory (LSTM) structure \cite{6795963}.
Vq-wav2vec \cite{baevski2019vq} utilizes a quantization module to extract discrete semantic units from unlabeled speech data and then uses the BERT to perform contextual modeling on the extracted units.
Wa2vec2.0 \cite{NEURIPS2020_92d1e1eb} employs a convolutional neural network (CNN) to extract local features from the raw waveform, which are then input to the BERT module to perform mask prediction by a contrastive loss.

To the best of our knowledge, there are few works focused on the robustness of self-supervised pre-trained models.
For example, it was shown in \cite{kawakami2020learning,riviere2020unsupervised} that a modified CPC pre-trained model can be transferred well across domains, and larger pre-training datasets lead the ASR model to be much more robust against domain shift.
The robust wav2vec2.0 proposed in \cite{hsu2021robust} reveals the impact of domain mismatch on the self-supervised speech representation learning.
The problem-agnostic speech encoder (PASE+) that was proposed in \cite{ravanelli2020multi} introduces online speech data augmentation modules for self-supervised learning and obtains a good performance in noisy environments.
Note that although in \cite{ravanelli2020multi} the multi-task self-supervised learning can obtain robust representations, the robustness against noise was not reported.
Apart from the requirement on the robustness in terms of the domain shift, the robustness of pre-trained models against noise should also be taken into account in order to evaluate the modeling effectiveness, particularly for practical noisy ASR applications.

In this work, we first investigate the noise robustness of the self-supervised pre-trained wav2vec2.0 model \cite{NEURIPS2020_92d1e1eb}.
For this, we use the same experimental data as in \cite{prasad2021investigation} for unsupervised pre-training and evaluate on the same noisy test set.
Experimental results show that pre-training on noisy data can obtain robust representations.
It is shown that the wav2vec2.0 model in noisy scenes can greatly improve the speech recognition performance of low signal-to-noise-ratio (SNR) speech, however, the performance significantly drops on the clean test set.
Therefore, in order to avoid the performance decrease on the clean test set, we then propose an enhanced wav2vec2.0 model.
In detail, during the pre-training phase, the noisy speech and clean speech are sent into a shared feature encoder.
The noisy feature is input to the transformer encoder, while the clean feature is fed to the vector-quantization (VQ) module, which provides clean training targets for the transformer encoder.
The proposed model is evaluated under different noisy conditions, which achieves a much better performance on noisy data at the cost of a tiny performance sacrifice on the clean test set, resulting in a better robustness compared to the original wav2vec2.0 as well as existing pre-training models.
The rest of this paper is arranged as follows. Section \ref{sec:method} introduces the classic wav2vec2.0 model and the proposed enhanced wav2vec2.0 model. Section \ref{sec:experiment} presents the data and model configurations. Results are shown in Section \ref{sec:result}, followed by concluding remarks in Section \ref{sec:conclusion}.
\vspace{-0.15cm}
\section{METHOD}
\vspace{-0.15cm}
\label{sec:method}

\subsection{Wav2vec2.0}
To guide the reader, we briefly introduce the wav2vec2.0 model in this section. In the wav2vec2.0, the raw waveform is input into a stack of CNN layers to obtain local features. A certain proportion of local features are masked and then sent to a contextual transformer network for predicting the masked features using the contextual information. Finally, the contrastive loss between the predicted features and the quantized features of the real frames is calculated. In this way, the pre-trained model enables good representations, which are rather beneficial for downstream ASR tasks. For instance, in case of only using 10 minutes of transcribed speech for fine-tuning, it achieves a word error rate (WER) of 5.7\%/10.1\% on the clean/noisy test sets of LibriSpeech \cite{panayotov2015librispeech}. For more details about the wav2vec2.0 model, we refer to \cite{NEURIPS2020_92d1e1eb}.

\begin{figure}[!t]
  \centering
\vspace{-0.5cm}
  \includegraphics[width=0.45\textwidth]{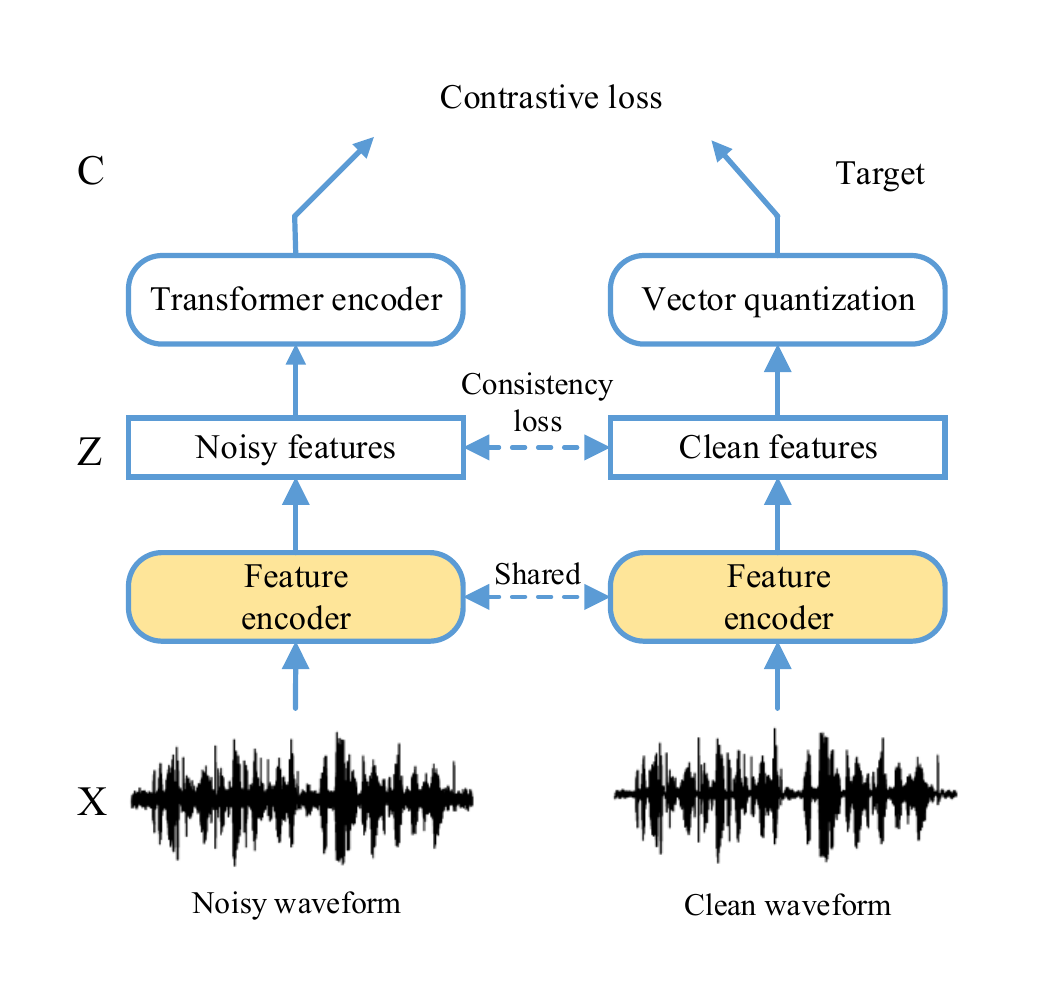}
  \vspace{-0.35cm}
  \caption{An illustration of the enhanced wav2vec2.0 model.}
  \vspace{-0.15cm}
  \label{fig:enhancedwav2vec2}
\end{figure}

\subsection{The proposed enhanced wav2vec2.0}
The proposed enhanced wav2vec2.0 model is on the basis of the classic wav2vec2.0~\cite{NEURIPS2020_92d1e1eb}, and the corresponding diagram is shown in Fig.~\ref{fig:enhancedwav2vec2}, which also consists of a feature encoder $f:X \mapsto Z$ and a transformer encoder $g: Z \mapsto C $.
The feature encoder consists of seven layers of convolutional network and the transformer encoder contains twelve transformer blocks.
Specifically, as shown in (\ref{eq1}), the shared feature encoder downsamples the input raw noisy waveform $X_{noisy}$ to the noisy features $Z_{noisy}$ and downsamples the input clean raw waveform $X_{clean}$ to the clean features $Z_{clean}$, i.e.,
\begin{equation}
  Z_{noisy} = f(X_{noisy}),\quad
  Z_{clean} = f(X_{clean}).
  \label{eq1}
\end{equation}
We mask a certain proportion of the noisy features $Z_{noisy}$, and replace it with a learnable vector at the masked position.
The transformer encoder then models high-level content from the input $Z_{noisy}$ into the noisy contextualized representations $C_{noisy}$, which is given by
\begin{equation}
  C_{noisy} = g(Z_{noisy}).
  \label{eq3}
\end{equation}
On the other hand, the corresponding clean features $Z_{clean}$ are discretized to $q_{clean}$ by a VQ module $Z \mapsto Q $, which are then used as clean targets in the contrastive objective, i.e.,
\begin{equation}
  q_{clean} = VQ(Z_{clean}).
  \label{eq4}
\end{equation}
The motivation of using clean features as the target originates from the expectation that the model can learn clean speech representations from noisy features. The involved  VQ module is implemented using the product quantization \cite{jegou2010product}.
Specifically, the VQ module first maps the clean features $Z_{clean}$ to logits $\mathbf{l} \in \mathbb{R}^{G \times V} $, where $G$ denotes the number of codebooks and $V$ denotes the number of entries in each codebook. The gumbel softmax function \cite{jang2016categorical} is then used to select discrete codebook entries in a fully differentiable way. As a result, for a given frame $Z_{clean_t}$ at time $t$ , we can select one entry from each codebook, concatenate the resulting vectors $e_1,...,e_G$ and apply a linear transformation to obtain $ q_{clean_t} $. The loss function can therefore be formulated as
\begin{equation}
  L = L_{m}+\alpha L_{d} + \beta L_{f} + \gamma L_c,
  \label{eq5}
\end{equation}
where
\begin{align}
  L_{m} &= -\log \frac{\exp({\rm sim}(C_{noisy_t},q_{clean_t})/\kappa)}{\sum_{\tilde{q} {\sim} Q_t}\exp({\rm sim}(C_{noisy_t},\tilde{q})/\kappa)},
  \label{eq6} \\
  L_{d} &= \frac{1}{GV}\sum_{g=1}^{G}\sum_{v=1}^{V}\overline{p}_{g,v}\log \overline{p}_{g,v},
  \label{eq7}\\
  \overline{p}_{g,v}&=\frac{\exp(\overline{l}_{g,v}+n_{v})/\tau}{\sum_{k=1}^{V}\exp(\overline{l}_{g,k}+n_k)/\tau},
  \label{eq8} \\
  L_{c} &= \left\| Z_{{noisy}_t}-Z_{{clean}_t} \right\|_2,
  \label{eq9}
\end{align}
which applies to any time index $t$.
It is clear that the total loss function is the weighted summation over four terms depending on the parameters $\alpha$, $\beta$ and $\gamma$. In (\ref{eq5}), $L_{m}$ is the contrastive loss, which enables the model distinguish between the true quantized clean features $q_{clean_t}$ and a set of $K+1$ quantized candidate features $\tilde{q} \in Q_{t}$. The quantized candidate features $\tilde{q}$ contains $q_{clean_t}$ and $K$ distractors. The diversity loss $L_{d}$ aims to increase the use of quantized codebook features, and $L_{f}$ is an $\ell_2$ penalty over the outputs of the feature encoder.  In (\ref{eq6}), sim stands for the cosine similarity between two vectors and $\kappa$ is a temperature. In (\ref{eq7}), $\overline{p}_{g,v}$ represents the probability of choosing the $v$-th codebook entry for group $g$ across a batch of utterances, where $\tau$ is a temperature. In (\ref{eq8}), $\overline{l}_{g,v}$ stands for the average logits $\mathbf{l}$ across utterances in a batch. In order to ensure the consistency between the clean features and the noisy features corrupted by noise, we additionally introduce a consistency loss. As defined in (\ref{eq9}), the  consistency loss $L_c$ measures the Euclidean distance between noisy features $Z_{noisy}$ and clean features $Z_{clean}$. In practice, the weighting parameters are set empirically.

\begin{table*}[]
\caption{ The performance comparison of different methods on type-A noise test sets at different SNRs, where ``No" means that the pre-training step is not included. The pre-training and fine-tuning can be performed on different datasets.}
\label{tab:table0}
\centering
\renewcommand\arraystretch{1.2}
\scalebox{0.8}{
\begin{tabular}{c|l|l|cccccccccccccccc}
\hline
\multirow{3}{*}{\textbf{Method}}   & \multicolumn{1}{c|}{\multirow{3}{*}{\textbf{Pre-train}}} & \multicolumn{1}{c|}{\multirow{3}{*}{\textbf{Fine-tune}}} & \multicolumn{16}{c}{\textbf{WER under SNR (dB)}}                                                                                                                                                                                                                          \\ \cline{4-19}
                          & \multicolumn{1}{c|}{}                           & \multicolumn{1}{c|}{}                           & \multicolumn{5}{c|}{\textbf{Traffic}}                          & \multicolumn{5}{c|}{\textbf{Metro}}                            & \multicolumn{5}{c|}{\textbf{Car}}                              & \multicolumn{1}{l}{\multirow{2}{*}{\begin{tabular}[c]{@{}l@{}}\textbf{Clean}\\ \textbf{WER}\end{tabular}}} \\ \cline{4-18}
                          & \multicolumn{1}{c|}{}                           & \multicolumn{1}{c|}{}                           & \textbf{0}    & \textbf{5}    & \textbf{10}   & \textbf{15}   & \multicolumn{1}{c|}{\textbf{20}}   & \textbf{0}    & \textbf{5}    & \textbf{10}   & \textbf{15}   & \multicolumn{1}{c|}{\textbf{20}}   & \textbf{0}    & \textbf{5}    & \textbf{10}   & \textbf{15}   & \multicolumn{1}{c|}{\textbf{20}}   & \multicolumn{1}{l}{}                                                                     \\ \hline
Baseline \cite{prasad2021investigation}                    & No                                              & Clean                                           & 72.4 & 62.5 & 50.2 & 41.0 & \multicolumn{1}{c|}{33.6} & 68.4 & 54.4 & 46.4 & 34.9 & \multicolumn{1}{c|}{27.6} & 35.0 & 28.1 & 24.3 & 21.7 & \multicolumn{1}{c|}{16.7} & \textbf{10.3}                                                                                     \\ \cline{1-3}
DEMUCS \cite{prasad2021investigation}                     & FreeSound                                       & FreeSound                                       & 38.2 & 30.3 & 25.3 & 20.6 & \multicolumn{1}{c|}{17.9} & 35.6 & 24.9 & 22.6 & 17.1 & \multicolumn{1}{c|}{15.9} & 20.5 & 18.1 & 14.6 & 13.8 & \multicolumn{1}{c|}{13.1} & 10.9                                                                                     \\ \cline{1-3}
AvT \cite{prasad2021investigation}                        & No                                              & FreeSound                                       & 40.7 & 32.5 & 26.3 & 21.4 & \multicolumn{1}{c|}{18.5} & 36.1 & 26.5 & 22.6 & 18.4 & \multicolumn{1}{c|}{17.8} & 21.8 & 18.9 & 16.8 & 16.0 & \multicolumn{1}{c|}{15.3} & 13.1                                                                                     \\ \hline
\multirow{5}{*}{Wav2vec2.0} & No                                              & Clean                                           & 71.0 & 57.8 & 48.6 & 40.5 & \multicolumn{1}{c|}{33.3} & 66.5 & 54.4 & 48.6 & 35.2 & \multicolumn{1}{c|}{28.2} & 34.1 & 27.1 & 22.8 & 19.2 & \multicolumn{1}{c|}{14.8} & 11.0                                                                                     \\ \cline{2-3}
                          & No                                              & FreeSound                                       & 52.3 & 44.8 & 38.9 & 34.6 & \multicolumn{1}{c|}{31.2} & 49.9 & 41.1 & 36.3 & 31.6 & \multicolumn{1}{c|}{30.0} & 34.7 & 30.3 & 29.5 & 28.4 & \multicolumn{1}{c|}{26.8} & 25.0                                                                                     \\ \cline{2-3}
                          & Clean                                           & FreeSound              & 42.8 & 34.2 & 26.7 & 23.0 & \multicolumn{1}{c|}{19.4} & 39.2 & 31.0 & 25.9 & 21.4 & \multicolumn{1}{c|}{19.7} & 23.0 & 19.0 & 17.6 & 16.2 & \multicolumn{1}{c|}{15.4} & 14.0                                                                                     \\ \cline{2-3}
                          & FreeSound                                       & FreeSound                                       & 34.6 & 27.9 & 23.6 & 18.9 & \multicolumn{1}{c|}{17.6} & 32.4 & 24.8 & 20.7 & 17.4 & \multicolumn{1}{c|}{17.1} & 19.4 & 17.0 & 15.5 & 14.7 & \multicolumn{1}{c|}{14.6} & 13.5                                                                                     \\ \cline{2-3}
                          & NoiseX-92                                         & FreeSound                                       & 35.4 & 28.4 & 24.3 & 21.0 & \multicolumn{1}{c|}{18.8} & 35.2 & 26.3 & 23.4 & 19.8 & \multicolumn{1}{c|}{18.5} & 21.2 & 18.2 & 17.6 & 17.1 & \multicolumn{1}{c|}{16.3} & 16.1                                                                                     \\ \hline
\multirow{2}{*}{Ours}     & FreeSound                                       & FreeSound                                       & \textbf{31.0} & \textbf{22.3} & \textbf{20.0} & \textbf{16.5} & \multicolumn{1}{c|}{\textbf{14.9}} & \textbf{29.0} & \textbf{21.9} & \textbf{18.1} & \textbf{15.8} & \multicolumn{1}{c|}{\textbf{14.4}} & \textbf{17.6} & \textbf{15.6} & \textbf{14.4} & \textbf{13.7} & \multicolumn{1}{c|}{\textbf{13.1}} & 12.3                                                                                     \\ \cline{2-3}
                          & NoiseX-92                                         & FreeSound                                       & 31.2 & 23.6 & 21.0 & 17.7 & \multicolumn{1}{c|}{16.8} & 29.1 & 23.0 & 19.7 & 17.1 & \multicolumn{1}{c|}{15.9} & 17.9 & 16.4 & 15.7 & 15.3 & \multicolumn{1}{c|}{14.6} & 14.3                                                                                     \\ \hline
\end{tabular}}
\end{table*}

\section{EXPERIMENTal setup}
\label{sec:experiment}
\subsection{Data description}

In order to facilitate a fair comparison with existing approaches, the data used in the experiments keep exactly the same as that in \cite{prasad2021investigation}. In this work, we utilize the Librispeech \cite{panayotov2015librispeech} train-clean-100 subset as the clean speech training set and the dev-clean subset as the validation set. The noise data used in the experiment come from FreeSound \cite{font2013freesound} and the sampling frequency of these noise data is 16 kHz. The noise type is divided into two categories, i.e.,  A and B. The type-A noise is relatively stationary, including `Car', `Metro' and `Traffic' noise, and the type-B noise is relatively non-stationary, including `Babble', `Airport/Station', `Cafe' and `AC/Vacuum' noise. Each type of noise has 10 and 8 different audio streams in the training set and the test set, respectively. The length of the noise data set is around 2 hours in total. In the process of model training and validation, we randomly select noise samples and mix with the clean speech at different SNRs $\in \lbrace 0,5,10,15,20,25 \rbrace $ dB to generate noisy data. For the test set, we first randomly select 120 clean speech from the test-clean subset in the Librispeech and then mix with noises at different SNRs $\in \lbrace 0, 5, 10, 15, 20 \rbrace $ dB to synthetize 4200 noisy test data. Note that the construction of the noisy data set keeps the same as \cite{nicolson2020deep}. The noise data and noisy test sets can be downloaded from the website\footnote{https://github.com/archiki/Robust-E2E-ASR}.
In addition, we also utilize the NoiseX-92 \cite{varga1993assessment} noise dataset at the pre-training stage  for supplementary experiments.

\subsection{Self-supervised pre-training}

In this work, the pre-training model is implemented using the fairseq toolkit \cite{ott2019fairseq}, which mainly includes a feature encoder module, a transformer encoder module and a VQ module. In detail, the feature encoder consists of seven convolutional layers and the channel number of the convolution module is 512. The stride and kernel sizes of the convolution module are (5, 2, 2, 2, 2, 2, 2),  (10, 3, 3, 3, 3, 2, 2), respectively. Therefore, the frame shift of the output $Z$ of the feature encoder is 20 ms and its receptive field is 25 ms. Both the clean speech and noisy speech are input to a shared feature encoder module to obtain speech features. For the transformer encoder module, we utilize 12  transformer encoder layers and each layer contains a self-attention module and a feed forward module. The dimension of the self-attention module is 512, and 8 heads are utilized. The dimension of the feed forward module is 512, and the inner dimension is 2048. For the VQ module, we set $G$ = 2 and $V$ = 320, and the dimension of each entry is 128. The model size including all parameters is around 45M. For masking, we sample at all time steps with a probability of $p$ = 0.065, and mask the subsequent $M$ = 10 time steps. For the loss function, the temperature $\kappa$ is set to be 0.1, and $\tau$ is annealed from 2 to 0.5 with a coefficient of 0.999995 in terms of iterations. The parameters $\alpha$, $\beta$ and $\gamma$ are set to be  0.1, 10 and 1, respectively. The number of distractors $K$ equals 100. The pre-training model utilizes the Adam optimizer \cite{kingma2014adam}. During the first 8\% of all epochs, the learning rate warms up to $5\times 10^{-4}$ and then decays linearly.

We randomly select noise samples and add them to clean speech at different SNRs $\in \lbrace 0,5,10,15,20,25 \rbrace $ dB to obtain the corresponding noisy speech. Then the noisy and clean speech are fed into the model, respectively.  The pre-training model is trained using 6 Tesla-V100-32G GPUs with 500 epochs, and the total training time is about 60 hours.

\subsection{Fine-tuning with labeled data}

After the pre-training is completed, we remove the VQ module on the basis of the pre-trained model, add an additional linear layer on the top of the transformer encoder, and then fine-tune the model using the labeled data. The modeling unit of the model has 30 characters, including 26 letters and 4 special symbols. The model is optimized using the connectionist temporal classification (CTC) loss function \cite{graves2006connectionist}. During fine-tuning, we use the noisy speech at different SNRs to fine-tune the model, and the generation of noisy data is the same as that in the pre-training phase. After the model fine-tuning, we decode on clean test sets and noisy test sets without any language model and then calculate the corresponding WER.

\begin{table*}[]
\caption{ The performance comparison of different methods on type-B noise test sets at different SNRs, where for brevity of presentation the results for the case of SNR = 0 dB are omitted (which does not affect the analysis of results).}
\label{tab:table2}
\centering
\renewcommand\arraystretch{1.2}
\scalebox{0.82}{
\begin{tabular}{c|l|l|cccccccccccccccc}
\hline
\multirow{3}{*}{\textbf{Method}}   & \multicolumn{1}{c|}{\multirow{3}{*}{\textbf{Pre-train}}} & \multicolumn{1}{c|}{\multirow{3}{*}{\textbf{Fine-tune}}} & \multicolumn{16}{c}{\textbf{WER under SNR (dB)}}                                                                                                                                                                   \\ \cline{4-19}
                          & \multicolumn{1}{c|}{}                           & \multicolumn{1}{c|}{}                           & \multicolumn{4}{c|}{\textbf{Babble}}                            & \multicolumn{4}{c|}{\textbf{Airport/Station}}                  & \multicolumn{4}{c|}{\textbf{AC/Vacuum}}                        & \multicolumn{4}{c}{\textbf{Cafe}}         \\ \cline{4-19}
                          & \multicolumn{1}{c|}{}                           & \multicolumn{1}{c|}{}                                & \textbf{5}    & \textbf{10}   & \textbf{15}   & \multicolumn{1}{c|}{\textbf{20}}      & \textbf{5}    & \textbf{10}   & \textbf{15}   & \multicolumn{1}{c|}{\textbf{20}}       & \textbf{5}    & \textbf{10}   & \textbf{15}   & \multicolumn{1}{c|}{\textbf{20}}       & \textbf{5}    & \textbf{10}   & \textbf{15}   & \textbf{20}   \\ \hline
Baseline \cite{prasad2021investigation}                 & No                                              & Clean                                            & 98.3 & 91.3 & 79.7 & \multicolumn{1}{c|}{65.0}  & 84.1 & 73.7 & 60.6 & \multicolumn{1}{c|}{50.0}  & 83.1 & 71.5 & 59.5 & \multicolumn{1}{c|}{45.8}  & 72.7 & 59.5 & 44.3 & 33.4 \\ \cline{1-3}
DEMUCS \cite{prasad2021investigation}                   & FreeSound                                       & FreeSound                                         & 58.0 & 41.8 & 32.3 & \multicolumn{1}{c|}{25.4}  & 45.5 & 33.7 & 25.6 & \multicolumn{1}{c|}{21.5}  & 45.4 & 34.2 & 28.1 & \multicolumn{1}{c|}{22.8}  & 31.6 & 27.4 & 20.3 & 16.9 \\ \cline{1-3}
AvT \cite{prasad2021investigation}                      & No                                              & FreeSound                                         & 55.1 & 39.5 & 31.1 & \multicolumn{1}{c|}{24.6}  & 43.3 & 33.4 & 25.2 & \multicolumn{1}{c|}{20.9}  & 40.8 & 33.4 & 29.3 & \multicolumn{1}{c|}{23.2}  & 32.0 & 26.3 & 21.4 & 18.5 \\ \hline
\multirow{5}{*}{Wav2vec2.0} & No                                              & Clean                                             & 93.1 & 84.9 & 73.7 & \multicolumn{1}{c|}{58.0}  & 80.6 & 72.7 & 59.7 & \multicolumn{1}{c|}{48.8}  & 79.6 & 69.6 & 56.5 & \multicolumn{1}{c|}{42.4}  & 68.9 & 58.1 & 43.7 & 34.8 \\ \cline{2-3}
                          & No                                              & FreeSound                                         & 70.8 & 58.6 & 47.6 & \multicolumn{1}{c|}{39.8}  & 59.5 & 49.4 & 40.6 & \multicolumn{1}{c|}{35.4}  & 56.5 & 48.4 & 42.4 & \multicolumn{1}{c|}{35.5}  & 49.6 & 42.7 & 34.6 & 31.8 \\ \cline{2-3}
                          & Clean                                           & FreeSound                                         & 58.3 & 45.0 & 35.1 & \multicolumn{1}{c|}{27.6}  & 48.1 & 37.3 & 28.8 & \multicolumn{1}{c|}{24.7}  & 44.9 & 36.1 & 29.2 & \multicolumn{1}{c|}{24.5}  & 36.4 & 29.1 & 23.9 & 19.2 \\ \cline{2-3}
                          & FreeSound                                       & FreeSound                                         & 47.4 & 35.9 & 29.0 & \multicolumn{1}{c|}{23.5}  & 39.6 & 29.4 & 23.7 & \multicolumn{1}{c|}{20.2}  & 40.9 & 32.4 & 26.8 & \multicolumn{1}{c|}{21.3}  & 28.0 & 24.6 & 19.7 & 17.0 \\ \cline{2-3}
                          & NoiseX-92                                       & FreeSound                                        & 52.1 & 39.7 & 30.6 & \multicolumn{1}{c|}{26.6}  & 40.7 & 30.8 & 25.3 & \multicolumn{1}{c|}{22.2}  & 45.8 & 37.0 & 30.7 & \multicolumn{1}{c|}{24.7}  & 30.2 & 25.3 & 22.2 & 18.6 \\ \hline
\multirow{2}{*}{Ours}     & FreeSound                                       & FreeSound                                         & \textbf{41.0} & \textbf{30.2} & \textbf{25.1} & \multicolumn{1}{c|}{\textbf{19.2}}  & \textbf{33.4} & \textbf{24.4} & \textbf{19.7} & \multicolumn{1}{c|}{\textbf{16.9}}  & \textbf{32.4} & \textbf{25.6} & \textbf{21.5} & \multicolumn{1}{c|}{\textbf{17.6}}  & \textbf{24.7} & \textbf{21.4} & \textbf{17.7} & \textbf{15.1} \\ \cline{2-3}
                          & NoiseX-92                                       & FreeSound                                         & 46.2 & 37.2 & 27.0 & \multicolumn{1}{c|}{22.3}  & 34.9 & 27.6 & 22.7 & \multicolumn{1}{c|}{19.3}  & 36.5 & 29.6 & 23.0 & \multicolumn{1}{c|}{19.0}  & 24.9 & 21.8 & 18.2 & 16.6 \\ \hline
\end{tabular}}
\end{table*}

\section{Experimental Results}
\label{sec:result}

\textbf{Comparison methods: }The Baseline in \cite{prasad2021investigation} utilizes the Deepspeech2 model  \cite{amodei2016deep} for training on the Librispeech train-clean-100 dataset with a CTC objective function and evaluates on different test sets. DEMUCS that was originally proposed in \cite{defossez20_interspeech} is a typical waveform-to-waveform model based on the introduction of a front-end speech denoising module \cite{prasad2021investigation}. The AvT method utilized in \cite{prasad2021investigation} introduces a gradient reversal layer (GRL) \cite{ganin2015unsupervised} in prior to the model classification layer, such that the learned speech representations can be noise-invariant. For specific implementation about DEMUCS and AvT, we refer to \cite{prasad2021investigation}. In addition, the proposed enhanced wav2vec2.0 model will also be compared with the original wav2vec2.0 method~\cite{NEURIPS2020_92d1e1eb}. Note that different combinations of pre-training and fine-tuning settings will be considered in experiments.

Table~\ref{tab:table0} and Table~\ref{tab:table2} show the ASR performances in terms of WER of the aforementioned approaches using the type-A (relatively stationary) and type-B (non-stationary) noises under different SNR conditions, respectively. From Table~\ref{tab:table0}, it can be seen that although the structure of the proposed model is different from \cite{prasad2021investigation}, the proposed baseline system (i.e., wav2vec2.0 no pre-train clean fine-tune) achieves a comparable performance as compared to the baseline in \cite{prasad2021investigation}.  For wav2vec2.0, comparing `no pre-train clean fine-tune' and `no pre-train FreeSound fine-tune' it is clear that fine-tuning on noisy datasets can  improve the ASR performance under most noise conditions, that is, the noise robustness can be improved. As the combination of `clean pre-train FreeSound fine-tune' obtains a much better performance than `no pre-train FreeSound fine-tune' in both noisy and clean environments, the inclusion of a pre-training phase is rather beneficial for the robustness of ASR models. Comparing `clean pre-train FreeSound fine-tune' and `FreeSound pre-train FreeSound fine-tune' (the latter performs better), we find that the wav2vec2.0 model can still learn a robust speech representation under noisy scenarios.  Compared to DEMUCS or AvT, although wav2vec2.0 (i.e., FreeSound pre-train FreeSound fine-tune) can improve the performance on the test set under various noisy conditions, the performance on the clean test set drops significantly. Similar conclusions can be draw from Table~\ref{tab:table2} as well.

In order to verify whether the wav2vec2.0 model is robust to noise types, we use the NoiseX-92 noise dataset to dynamically add noise to the train-clean-100 subset to obtain noisy dataset for pre-training and then perform fine-tuning on noisy data. From Table~\ref{tab:table0} and Table~\ref{tab:table2}, we can see that the choice of `NoiseX-92 pre-train FreedSound fine-tune' for wav2vec2.0 is better than the  `no pre-train FreedSound fine-tune' counterpart, indicating that the representations obtained by pre-training on other types of noisy data still have a good robustness. However, as the choice of `FreeSound pre-train FreeSound fine-tune'  leads to a decrease in WER compared to `NoiseX-92 pre-train FreeSound fine-tune', the data source for pre-training and fine-tuning affects the performance of wav2vec2.0. That is, the noise data for pre-training and fine-tuning originating from different domains might degrade the ASR performance.

Furthermore, from Table~\ref{tab:table0} and Table~\ref{tab:table2} we can see that the combination of `FreedSound pre-train FreedSound fine-tune' for the proposed method  is better than the same choice for wav2vec2.0 under both noisy and clean conditions. Using the clean speech as the pre-training targets can improve the performance on the noisy test set, and it is also ensured that the performance on the clean test set is not significantly degraded. In addition, the proposed method with `NoiseX-92 pre-train FreedSound fine-tune' is better than the wav2vec2.0 counterpart, indicating that a better robustness against different noise types is obtained. Besides, although the proposed enhanced wav2vec2.0 approach works slightly worse than DEMUCS, the performance under noisy conditions is much better, which are more common to happen in practice.

Finally, we compare the output representations of different models after fine-tuning on the noisy test set using the cosine similarity criterion.
The cosine similarity measures the similarity of vector $\mathbf{a}$ and vector $\mathbf{b}$, which is defined as $similarity(\mathbf{a}, \mathbf{b})=\mathbf{a}^T\mathbf{b}/\left\|\mathbf{a} \|\| \mathbf{b} \right\|$, and its value range is [0, 1].
It is clear that the larger the cosine similarity, the more similar between two vectors. Therefore, we can regard the clean baseline as the target and compare the cosine similarity between the output features of the transformer encoder of each model and the clean baseline model.  Table~\ref{tab:table3} shows the cosine similarity under the traffic and babble noise conditions. 
It can be seen from  Table~\ref{tab:table3} that the greater the SNR of the test set, the greater the similarity between the learned representation and the clean version. We can see that the cosine similarity obtained by the proposed enhanced wav2vec2.0 model is largest, which indicates that the feature of the enhanced wav2vec2.0 model output is the closest to the true clean feature.
That is, even under noisy conditions the proposed approach enables a higher fidelity speech representation with respect to the clean speech compared to existing methods, which is rather helpful for the subsequent ASR.

\begin{table}[]
\caption{ The cosine similarity between the transformer encoder output features of different models and the clean baseline model under the traffic and babble noise conditions.}
\label{tab:table3}
\renewcommand\arraystretch{1.2}
\centering
\scalebox{0.8}{
\begin{tabular}{l|ccccc}
\hline
\multicolumn{1}{c|}{\multirow{2}{*}{\textbf{Model}}}                                                    & \multicolumn{5}{c}{\textbf{ Cosine similarity}}                                                                                \\ \cline{2-6}
\multicolumn{1}{c|}{}                                                                          & \textbf{0 dB}                 & \textbf{5 dB}                 & \textbf{10 dB}                & \textbf{15 dB}                & \textbf{20 dB}                \\ \hline
\multicolumn{1}{c|}{\textbf{Traffic} (type-A noise)}                                                                   & \multicolumn{1}{l}{} & \multicolumn{1}{l}{} & \multicolumn{1}{l}{} & \multicolumn{1}{l}{} & \multicolumn{1}{l}{} \\
\begin{tabular}[c]{@{}l@{}}Wav2vec2.0 (no pre-train\\ FreeSound fine-tune)\end{tabular}        & 0.599               & 0.626               & 0.645               & 0.658               & 0.669               \\ \cline{1-1}
\begin{tabular}[c]{@{}l@{}}Wav2vec2.0 (FreeSound pre-train\\ FreeSound fine-tune)\end{tabular} & 0.732               & 0.752               & 0.767               & 0.776               & 0.783               \\ \cline{1-1}
\begin{tabular}[c]{@{}l@{}}Ours (FreeSound pre-train\\ FreeSound fine-tune)\end{tabular}       & \textbf{0.765}               & \textbf{0.781}               & \textbf{0.789}               & \textbf{0.795}               & \textbf{0.799}               \\ \hline\hline
\multicolumn{1}{c|}{\textbf{Babble} (type-B noise)}                                                                    &                      &                      &                      &                      &                      \\
\begin{tabular}[c]{@{}l@{}}Wav2vec2.0 (no pre-train\\ FreeSound fine-tune)\end{tabular}        & 0.510               & 0.552               & 0.589               & 0.622               & 0.644               \\ \cline{1-1}
\begin{tabular}[c]{@{}l@{}}Wav2vec2.0 (FreeSound pre-train\\ FreeSound fine-tune)\end{tabular} & 0.655               & 0.699               & 0.730               & 0.753               & 0.767               \\ \cline{1-1}
\begin{tabular}[c]{@{}l@{}}Ours (FreeSound pre-train\\ FreeSound fine-tune)\end{tabular}       & \textbf{0.709}               & \textbf{0.741}               & \textbf{0.765}               & \textbf{0.781}               & \textbf{0.791}               \\ \hline
\end{tabular}}
\end{table}

\vspace{-0.1cm}
\section{CONCLUSION}
\label{sec:conclusion}

In this paper, we investigated the robustness of  the self-supervised pre-training model, i.e., wav2vec2.0. It was shown that pre-training on noisy data can obtain good representations. The wav2vec2.0 can improve the ASR performance on the noisy test set, but the performance on the clean test set drops. By taking the clean speech as the training target for pre-training, the proposed enhanced wav2vec2.0 model can learn a better speech representation, which thus improves the ASR performance and avoids an obvious  performance degradation on the clean test. In addition, it was shown that the proposed method is also robust to different types of noise.
 We found that the noise robustness of ASR models is related to the fidelity (or SNR) of learned speech representations, and the proposed method can be interpreted as a representation enhancement, which improves the fidelity of the target speech in principal.


\bibliographystyle{IEEEbib}
\bibliography{strings,refs}

\end{document}